**Title:** Mathematical and Computational Nuclear Oncology: Toward Optimized Radiopharmaceutical Therapy via Digital Twins

**Authors:** Marc Ryhiner MSc [1] (corresponding author), Yangmeihui Song MD PhD[2], Babak Saboury MD PhD[3], Gerhard Glatting PhD[4], Arman Rahmim PhD DABSNM[5, 6], Kuangyu Shi PhD[1]

**Affiliation:**

1. Department of Nuclear Medicine, Inselspital, University of Bern, Bern, Switzerland
2. Department of Nuclear Medicine, Union Hospital, Huazhong University of Science and Technology, Wuhan, China
3. Institute of Nuclear Medicine, Bethesda, USA
4. Medical Radiation Physics, Department of Nuclear Medicine, Ulm University, Ulm, Germany
5. Departments of Radiology and Physics, University of British Columbia, Vancouver, Canada
6. Integrative Oncology, BC Cancer Research Institute, 675 West 10th Avenue, Vancouver, BC V5Z 1L3, Canada

**Email:**

- Marc Ryhiner: marc.ryhiner@unibe.ch
- Yangmeihui Song: songymh@gmail.com
- Babak Saboury: babaksaboury@gmail.com
- Gerhard Glatting: gerhard.glatting@uni-ulm.de
- Arman Rahmim: arman.rahmim@ubc.ca
- Kuangyu Shi: kuangyu.shi@unibe.ch

**Synopsis:** This article presents the general framework of theranostic digital twins (TDTs) in computational nuclear medicine, designed to support clinical decision-making and improve cancer patient prognosis through personalized radiopharmaceutical therapies (RPTs). It outlines potential clinical applications of TDTs and proposes a roadmap for successful implementation. Additionally, the chapter provides a conceptual overview of the current state of the art in the mathematical and computational modeling of RPTs, highlighting key challenges and the strategies being pursued to address them.

**Key Points:**

- Theranostic digital twins (TDTs) aim to improve cancer patient outcomes by enabling personalized therapeutic planning based on diagnostic imaging, particularly for radiopharmaceutical treatments.
- While several model components of TDTs have already been developed, upcoming efforts focus on integrating these models to support clinical translation.
- Current research in the field is increasingly focused on developing targeted solutions that capture the complex interactions between radiopharmaceutical therapies and human pathophysiology.

## 1. Introduction

Radiopharmaceutical therapies (RPTs) are emerging treatments for metastatic cancer, delivering ionizing radiation with high specificity to tumor tissue. A new renaissance in therapy development is being driven by improved biological understanding and technological advances such as artificial intelligence (AI) integration[1]. While RPTs have already shown promise not only in clinical trials[2,3], but also in routine medical practice[4], the current one–size–fits–all strategy almost often fails to achieve complete remission[5], possibly also because it does not address patient–specific tumor biology and resistance mechanisms[6]. Additionally, literature shows that fixed injected activities result in a wide variability in delivered doses, and that standardization of absorbed dose delivery would require personalization of administered activity[7]. Theranostic digital twins (TDTs) for RPTs represent a precision medicine approach designed to tailor treatment schedules to individual patients, thereby enhancing tumor response and therapeutic efficacy[8]. Personalized optimization of treatment parameters, such as cycle number and frequency, through TDTs could significantly improve patient survival.

Digital twins are virtual system replicas that comprise the relevant parameters of a biological system for describing a considered process[9], and they have significant potential for being applied to theranostics[10,11]. In this context, TDTs aim to model the complex interplay between tumor characteristics and treatment response[12]. Extended TDTs, beyond sole RPTs, incorporate additional modalities such as external beam radiotherapy (EBRT) or pharmacological agents and are further refined when guided by tumor–specific genomic features[13-15].

A TDT in RPTs is fundamentally a mathematical and computational model of both (i) pharmacokinetics and (ii) pharmacodynamics: namely what the body does to the drug and what the drug does to the body, respectively[16]. In RPTs, the former can be captured by realistic pharmacokinetic models, and the latter by radiation biology models that can quantitatively predict the patient's biological response. Constructing such a model poses challenges, particularly due to the patient–specific delivery of radiopharmaceuticals, as well as heterogeneity of absorbed dose at multiple physiological levels, and the mapping between absorbed dose and cell survival. The latter addresses the applicability of traditional models such as the linear-quadratic (LQ) model in the RPTs context[17]. This chapter presents the foundational concepts, mathematical and computational strategies, and biological principles needed to develop effective TDTs for optimizing cancer treatment.

## 2. RPTs Digital Twins for Personalized Cancer Treatment in Clinics

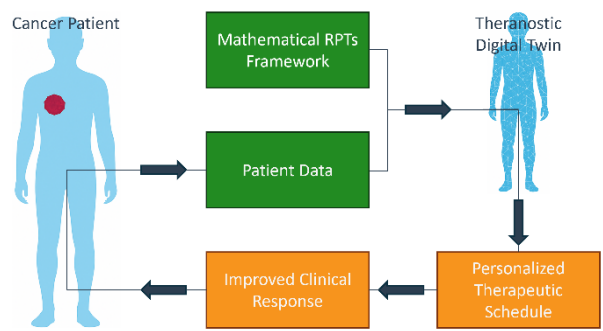

**Figure 1** Iterative interplay between the cancer patient and the theranostic digital twin (TDT). Patient–specific data are integrated into a mathematical framework to generate a digital twin capable of optimizing cancer treatment protocols. Clinical response data are then used to refine the twin, enhancing its predictive accuracy.

To address the lack of individualized RPT protocols, TDT simulations offer a compelling solution for tailoring therapeutic schedules to each patient, thereby optimizing the dose delivered to eliminate the lesions while sparing healthy tissues[18]. As virtual representations of biological systems, TDTs can be used to predict therapeutic outcomes and optimize decisions for improved efficacy[19].

The field of computational nuclear oncology involves the application of mathematical and computational models to simulate and analyze the behavior of radiopharmaceuticals within the body, particularly in the context of theranostic approaches[20]. Advanced modeling is essential in nuclear medicine to capture the complex interplay of biological and physical processes that determine the outcomes of RPTs[21]. These tools enable the personalization of treatment strategies and support the exploration of novel therapeutic concepts that would be difficult to investigate through empirical methods alone. In a computational nuclear oncology approach, TDTs integrate anatomical, physiological, and pathophysiological patient data into a framework representing RPTs-specific pharmacokinetics and pharmacodynamics[22] (Fig. 1). Treatment personalization is based on tumor characteristics, including observable parameters (lesion size, location, and tissue type), derived quantities (immune profile and receptor density), and modeled variables (vascularization, microdosimetry, and absorbed dose–response phenotype)[23]. The goal is to optimize the trade-off between maximizing tumor shrinkage and minimizing healthy tissue exposure by adjusting the radionuclide type, injected activity, treatment cycle number and frequency, and adjunct therapies[24].

Beyond general frameworks, reduced models that are not personalized, and thus not digital twins, are often developed to address specific clinical questions. For example, *in silico* studies have investigated optimal combinations of radiopharmaceutical molar amounts and activities to maximize tumor control probability while minimizing toxicity to organs–at–risk[25,26]. Other models explore receptor-ligand kinetics to identify novel radiopharmaceutical candidates or desirable kinetic properties[27,28]. These focused models, when accurately parameterized, can yield clinically actionable insights.

Translating TDTs into clinical practice requires personalized, accurate estimates of both absorbed dose and the dose–response relationship, including reliability measures to support risk assessment[29]. Personalization relies on data from imaging, histology, and circulating tumor DNA (ctDNA) analysis[30,31].

Positron emission tomography (PET)/computed tomography (CT) scans for oncological applications can provide critical information on lesion location, morphology, and metabolic activity, offering insights into tumor immune status[32,33]. Simulated FDG uptake gradients within the tumor microenvironment can help to understand underlying vascularization, enabling indirect inference of perfusion heterogeneity[34]. Single photon emission computed tomography (SPECT)/CT using the therapeutic agent provides information on radiotracer biodistribution, organ and tumor absorbed doses, and early treatment response[35]. In the context of TDT calibration, intratumoral dose heterogeneity can further inform on receptor density variations within the tumor microenvironment[36].

Histological analyses yield detailed cellular and nuclear architecture, essential for accurate microdosimetry. Functional assessments of DNA damage response genes such as *TP53* and *BRCA1/2* are also critical, given their role in cell radiosensitivity[37]. Liquid biopsies allow for noninvasive genotyping of these mutations through ctDNA, with next-generation sequencing and bioinformatic pipelines enabling variant calling and interpretation against reference genomes and clinical databases[38]. For robust clinical integration of TDTs, key milestones include standardizing microdosimetry models[39], validating dose–response relationships[40], and unifying pharmacokinetic and pharmacodynamic models within a physiologically accurate framework[41] (Fig. 2).

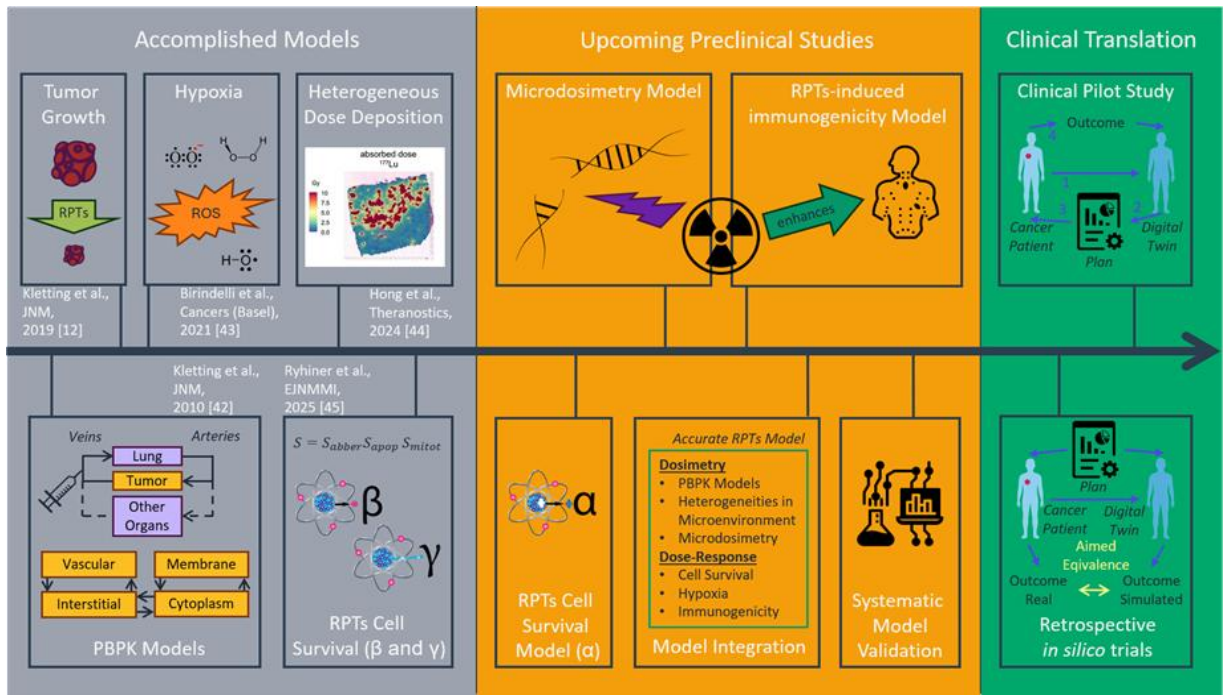

**Figure 2** Proposed roadmap for the clinical translation of theranostic digital twins (TDT) in radiopharmaceutical therapies (RPTs). Building on previously established models, the integration of frameworks for alpha therapy, microdosimetry, and RPTs–induced immunogenicity will precede clinical translation effort[12,42-45]. Prior to initiating clinical trials that assess the effectiveness of therapeutic schedules optimized by TDTs, extensive *in silico* trials will be conducted to aid the design of those clinical trials.

## 3. A General Framework for TDTs in RPTs

A comprehensive TDT for RPTs models every critical step from radiopharmaceutical injection to tumor regression. The foundation of such models lies in linking absorbed dose to cell survival probability, with a particular focus on DNA double-strand breaks (DSBs), the primary cytotoxic lesions induced by ionizing radiation[46]. Since DSBs occur in the nucleus, accurate modeling of nuclear dosimetry is central to predicting treatment outcomes[47].

The two major challenges in RPT modeling are: (1) accounting for spatial and temporal dose heterogeneities, and (2) elaborating patient–specific dose–response relationships.

### 3.1. Patient–Specific Dosimetry

Unlike EBRT, where dose delivery is relatively uniform and well-defined, RPTs dosimetry is inherently complex (Fig. 3). It depends on the spatial relationship between radionuclide decay and the target cell nucleus, as well as on cell and nucleus size, factors that vary across the tumor population[48].

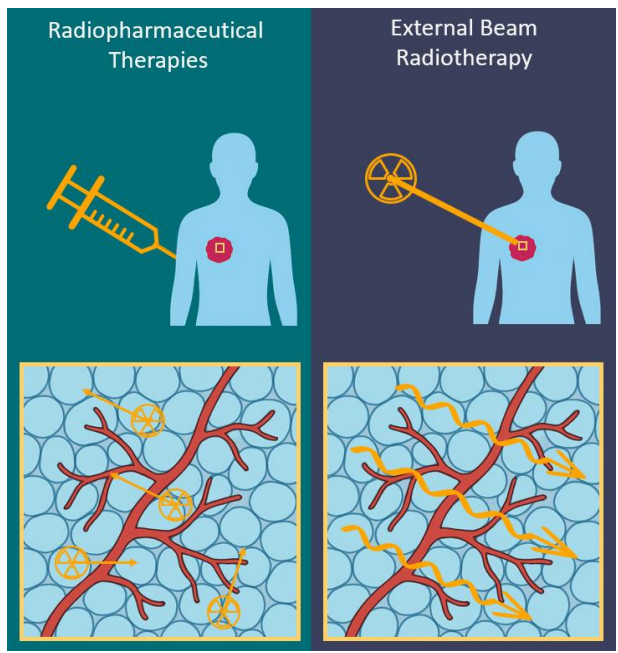

**Figure 3** Challenges in dosimetry modeling for radiopharmaceutical therapies (RPTs) compared to external beam radiatiotherapy (EBRT) arise primarily from multi–level heterogeneities. In EBRT, far-reaching gamma rays from an external source produce relatively homogeneous energy deposition across the targeted region, from the tissue to the cellular scale. In contrast, dose deposition in RPTs depends on the spatial distribution of decay events relative to individual cancer cells, as well as their geometries. Accurate RPT dosimetry for theranostic digital twins (TDTs) requires detailed pharmacokinetic modeling across organ, tissue, and cellular levels. Moreover, the range of the emitted radiation significantly influences the dose distribution. Temporal heterogeneities in dose delivery must also be considered when linking absorbed dose to biological response.

Two major advances in nuclear imaging significantly transformed dosimetry inference[49]. First, the integration of CT with PET enabled a shift from organ-level averages to voxel-level dosimetry[50]. Second, SPECT/CT improved accessibility and compatibility with a wider range of therapeutic isotopes, playing a crucial role in quantifying the spatial distribution of gamma-emitting radiopharmaceuticals[51]. On a finer scale, *in vitro* studies using radiometric counters can quantify radiopharmaceutical movement among the interstitial space, membrane, and cytoplasm to model ligand-receptor dynamics[52,53].

Physiologically-based pharmacokinetic (PBPK) models, when calibrated for patient-, compound-, and cell-specific parameters, can predict time-activity curves across biological levels[29,54,55]. For radiopharmaceuticals with decay chains (e.g., $^{225}$Ac, $^{223}$Ra, $^{212}$Pb), radionuclide generators and chelation stability are explicitly modeled to reflect realistic pharmacokinetics[56,57]. Target–mediated drug disposition (TMDD) modeling characterizes the interaction of radiopharmaceuticals with their biological targets, encompassing both receptor binding and internalization, and can be effectively integrated into PBPK frameworks[58].

To address tumor tissue heterogeneity, several strategies have emerged: diffusion/perfusion modeling based on histological data[30], spatial transcriptomics for mapping receptor and drug distribution[31], and deep learning approaches for voxel-wise dose prediction using pretherapeutic imaging[59]. The latter technique enhances the predictive power of TDTs by enabling localized and individualized treatment optimization.

For linking time-activity curves to absorbed dose rates, kernel–based dose-point computations that account for tissue heterogeneities are used, avoiding assumptions of uniform activity distribution that are particularly unreliable in low-range alpha therapies[60]. At the cell compartment level, the same link can be established using the MIRD (Medical Internal Radiation Dose) schema, which employs S values representing the dose rate per unit activity for each source–target compartment pair[61,62]. S values can be determined using Monte Carlo-based software tools, such as Geant4-DNA[63], GATE[64], or MIRDcell[65], under the assumption of defined cell geometries. Monte Carlo simulations use random sampling to model probabilistic processes[66]. This is particularly important in dosimetry, as they allow for highly accurate modeling of radiation interactions in the complex geometries and heterogeneous materials found in biological systems[67]. In practice, this means that the simulations can track the paths of individual radiation particles (photons, electrons, alpha particles) through tissue and record their energy deposition[68]. Obtaining accurate dose deposition is the aim in TDT dosimetry to lay the foundation for predicting therapeutic effects reliably[69].

### *3.2. Personalized Dose–Response Relationships*

The goal of the TDT for RPTs in dose–response modeling is to derive cell survival probabilities from time-resolved absorbed radiation dose rate curves[70] (Fig. 4). The LQ model (Eq. 1) quantifies cell survival ($S$) as a function of both linear ($\alpha D$) and quadratic components ($\beta D^2$) of the absorbed dose[71].

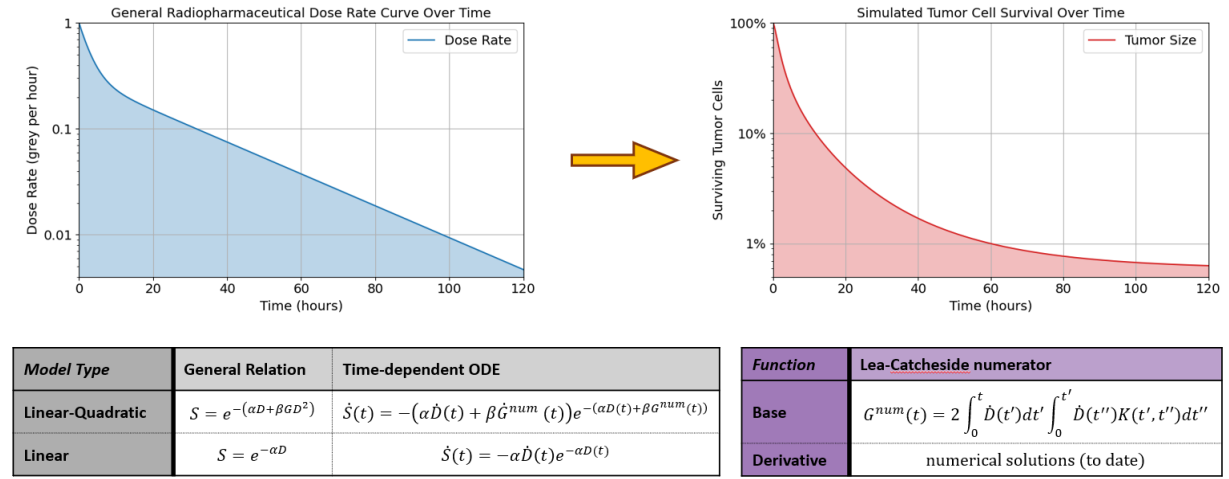

| Model Type | General Relation | Time-dependent ODE |
|---|---|---|
| Linear-Quadratic | $S = e^{-(\alpha D + \beta G D^2)}$ | $\dot{S}(t) = -(\alpha \dot{D}(t) + \beta \dot{G}^{num}(t))e^{-(\alpha D(t) + \beta G^{num}(t))}$ |
| Linear | $S = e^{-\alpha D}$ | $\dot{S}(t) = -\alpha \dot{D}(t)e^{-\alpha D(t)}$ |

| Function | Lea-Catcheside numerator |
|---|---|
| Base | $G^{num}(t) = 2\int_0^t \dot{D}(t')dt' \int_0^{t'} \dot{D}(t'')K(t',t'')dt''$ |
| Derivative | numerical solutions (to date) |

**Figure 4** The objective of quantitative dose–response research in radiopharmaceutical therapies (RPTs) is to map a given absorbed dose rate curve to corresponding probabilities of cell survival in both cancerous and healthy tissues. In this illustrative example, the dose rate curve represents a biexponentially declining trend, typically observed in practice due to the combined effects of radioactive decay and biological clearance. The resulting survival curve is simulated with a dose–response model[45]. Note: It is assumed that 1% of the energy released from decays within the tumor tissue is deposited in the cell nucle[72].

$$S = e^{-(\alpha D + \beta D^2)} \tag{1}$$

While it is widely used in clinical practice to predict outcomes of EBRT, it has limitations in the context of RPTs, where the radiation is delivered continuously[73]. Moreover, the various particle types used in RPTs lead to variability in the $\alpha$ and $\beta$ parameters, which depend on both radiation quality and cell type[74]. DNA repair mechanisms enable cells to better tolerate radiation when the dose is distributed over longer durations, as fewer DSBs occur simultaneously[75]. This reduces the probability of misrepair, which is more likely when DSBs accumulate concurrently[76]. The quadratic term in the LQ model accounts for the increased number of DSB ends and the associated higher likelihood of erroneous repair as damage accumulates[77]. Consequently, to accurately map the dose–response relationship in RPTs to cancer cell survival, it is insufficient to consider only the total absorbed dose. Instead, a time-resolved profile of the absorbed dose rate is required throughout the treatment course, for which the Lea-Catcheside time (Eq. 2) factor offers a way to extend the LQ model (Eq. 3) to account for temporal heterogeneity[78].

$$G = \frac{2}{D^2} \int_0^\tau \dot{D}(t) dt \int_0^t \dot{D}(t') K(t,t') dt' \tag{2}$$

$$S = e^{-(\alpha D + \beta G D^2)} \tag{3}$$

The Lea-Catcheside factor ($G$) reduces the quadratic term to account for the minimized chance of misinteraction between DSBs present at all possible time points ($t$ and $t'$) during the treatment duration ($\tau$) given the dose rate ($\dot{D}$) curves. The tissue repair kernel ($K$) can be modeled as a biexponential factor (Eq. 4) due to the biphasic kinetics of DSB clearance[79].

$$K(t,t') = p_1 e^{-\mu_1(t-t')} + p_2 e^{-\mu_2(t-t')} \tag{4}$$

Clinical applications often assume monoexponential DSB clearance with a single repair rate ($\lambda$) and sole contribution ($p = 1$), as this simplifies the Lea-Catcheside factor (Eq. 5) when monoexponential dose decay models are applied (valid when cell survival is evaluated at treatment end)[80]. The Hänscheid method, for example, uses a monoexponential form for the dose rate curve, with a decay rate ($\mu$) derived from the effective half-life[81].

$$G = \frac{\lambda}{\lambda + \mu} \tag{5}$$

Other RPT-specific models focus only on the linear component[82,83] (Eq. 6), which adequately represents RPT data since the limited number of concurrent DSBs rarely gives rise to a significant quadratic term ($\alpha D > \beta G D^2$)[84].

$$S = e^{-\alpha D} \tag{6}$$

Nonetheless, more sophisticated models are valuable when greater accuracy is needed or when accounting for additional factors, such as synergistic combination treatments[85]. A paper in this special issue elaborates on these considerations, including more advanced computational modeling in radiobiology[86].

In an alternative mechanistic framework, linking absorbed dose to cancer cell survival via DSBs is provided by the MEDRAS (Mechanistic DNA Repair and Survival Model) model, which was originally developed for external beam radiotherapy (EBRT)[87]. In MEDRAS, cells undergo necrosis or apoptosis based on the occurrence and misrepair of DSBs. The likelihood and consequences of DSB misrepair are determined by the fidelity of the DNA repair pathways and the presence of a homologous genome, both of which are dependent on the cell cycle phase[88]. Given the similar linear energy transfer characteristics of beta and gamma radiation, the parameters calibrated in MEDRAS can be directly applied to beta-emitting RPTs[89]. Modeling alpha therapy using MEDRAS appears infeasible not due to differences in spatial DSB distribution or the resulting variations in repair kinetics and fidelity, but rather because of the high proportion of unrepairable damage, which MEDRAS does not account for[90].

A recent study extended MEDRAS into an RPT-specific model[45]. In this framework, DNA damage and repair processes are driven by a time-resolved dose rate curve generated from a pharmacokinetic compartment model. Cell survival probability was then derived using a dynamic rate calibrated through MEDRAS, which incorporates both DNA damage levels and cell cycle information. While this adaptation effectively captures the continuous dose delivery of RPTs and models DSB induction and repair over time, it also has limitations. For instance, cellular senescence is not explicitly represented, misrepair eventually resulting in necrosis is time-wise equated with immediate cell death, and cell cycle delays induced by RPTs are not incorporated, despite the generally modeled cell cycle checkpoints. Tumor shrinkage is estimated through a combination of survival probability and cancer cell proliferation rates.

Quantitative Systems Pharmacology (QSP) models for immune–tumor interactions can simulate the complex dynamics between cancer cells and the immune system[91]. Since the immune system plays a critical role in counteracting tumor growth, TDT models should also incorporate immune–tumor interactions[92]. RPTs have been shown to enhance tumor immunogenicity, in part by increasing T-cell infiltration into the tumor microenvironment[93]. This could result from RPT-induced modifications to tumor architecture, which increase immune cell accessibility. However, this positive effect may be mitigated by hematologic toxicity associated with RPTs, which reduces leukocyte counts, potentially due to radiation crossfire effects on bone marrow[94]. TDTs that model these competing effects could help predict immune responses to RPT and exploit the patient's immune system as part of the therapeutic strategy. RPT efficacy is also influenced by tumor oxygenation[95], since reactive oxygen species mediate much of the associated DNA damage[96]. Accounting for hypoxia in TDTs may thus improve their predictive accuracy for treatment response[43]. The oxygen enhancement ratio (OER) quantifies how oxygen affects the biological effectiveness of radiation[97] (Eq. 7).

$$OER = \frac{D_1}{D_2} \tag{7}$$

Here, $D_1$ and $D_2$ are doses that produce the same biological effect under hypoxic and normoxic conditions, respectively.

### 3.3 Modeling Combination Therapies Involving RPTs

In addition to optimizing monotherapies, TDTs can inform combination treatment strategies, potentially leading to improved clinical outcomes. For example, combining RPTs with EBRT may be synergistic, particularly as EBRT has been reported to modulate PSMA (Prostate-Specific Membrane Antigen) receptor expression favorably[98], enhancing the effectiveness of subsequent RPTs. EBRT's strength lies in delivering high doses with maximal DSB clustering per unit dose, while RPTs, through continuous, low-level radiation, can evade cell cycle checkpoints by inducing sub-lethal DNA damage that escapes detection[99,100]. TDTs can help identify optimal dose combinations that maximize these complementary mechanisms.

Another promising strategy is combining alpha- and beta-emitting RPTs. Alpha particles have high energy and short path lengths, resulting in highly localized DNA damage with a high ratio of DSBs to single-strand breaks (SSBs)[101], reduced crossfire effects, and limited off-target toxicity[102]. These characteristics make alpha emitters more effective in treating micrometastases or small tumors, whereas beta emitters are better suited for larger tumors due to their longer range. TDTs can support individualized planning by identifying the most effective multi-isotope strategy for each patient[103].

Moreover, TDTs can guide combination regimens involving non-radioactive agents such as DNA damage response inhibitors or immune checkpoint inhibitors[104]. A recent study proposed a mechanistic model for the combination of RPTs with poly (ADP-ribose) polymerase (PARP) inhibitors (PARPi), incorporating the dose–response phenotype of homologous recombination deficiency (HRD)[45]. In HRD-positive cells, it was previously assumed that PARPi-induced inhibition of SSB repair leads to unrepaired lesions that convert into DSBs during S phase, a process resulting in synthetic lethality[105]. More recent findings suggest, however, that PARPi mainly stall replication forks, which can be restarted via homologous recombination[106]. Despite this, developed model successfully simulates experimental outcomes *in vitro*. TDTs are valuable tools for modeling the effects of different repair pathways under various treatment scenarios.

There is also preclinical evidence of synergy between [225]Ac and immune checkpoint inhibitors such as anti-PD1 antibodies, which can block the tumor's immune evasion mechanisms[107]. Future TDTs could incorporate such mechanisms to tailor combination therapies for patients based on their genomic and immunological profiles. Additionally, these models could support drug development by predicting the outcomes of novel treatment combinations through the simulation of relevant cellular pathways.

## 4. Computational Techniques for Building TDTs

A clinically viable TDT must accurately capture both therapeutic effects and patient responses to enable treatment planning[20]. Deterministic mathematical models use ordinary, partial, or delayed differential equations and yield the same outcomes for a given input at every time point.[108] They seem to show clear potential for TDTs, as the high number of tumor cells in cancer could ensure that the overall tumor exhibits relatively average behavior. However, probabilistic stochastic and AI-based models are increasingly used in digital medicine[109]. Machine learning, for instance, has been applied to predict dosimetry from limited time points and even pretherapeutically using diagnostic imaging[110,111]. Also, stochastic distributions appear particularly suitable for modeling biological heterogeneities and variabilities[112].

Most TDTs are currently centered on mechanistic models such as PBPK models, a trend likely to persist moving forward[113] (Fig. 5). These represent biological processes through systems of time-dependent differential equations based on established physiological laws[114]. To ensure real-world applicability by fitting unknown parameters to experimental data, the framework of verification, validation, and uncertainty quantification (VVUQ) is applied: verification confirms that the right equations are solved correctly, validation ensures equations accurately represent the real data, and uncertainty quantification provides solution variations propagated from input errors, which can be translated into confidence intervals and risk assessments to support decision-making[115]. External validation prevents overfitting to center-specific conditions and ensures the generalizability of a model[116]. Besides population-based confidence intervals, Bayesian methods also provide patient–specific confidence intervals, which are central to decision-making[117]. To ensure robust parameter estimation and prevent overfitting, structural and practical identifiability analyses are performed, particularly for highly parametrized models with limited clinical time-point data[118]. Structural identifiability verifies that each parameter is uniquely determinable from the model equations, while practical identifiability assesses whether parameters can be reliably calibrated with the available data[119]. To improve the robustness of models based on limited experimental data, various mathematical and computational techniques have been introduced to integrate a priori knowledge[120]. Analytical solutions to these systems are often impossible due to complexity, e.g., non–linearity; hence, numerical simulations and simplified approximations (e.g., steady-state assumptions or series expansions) are commonly employed[121].

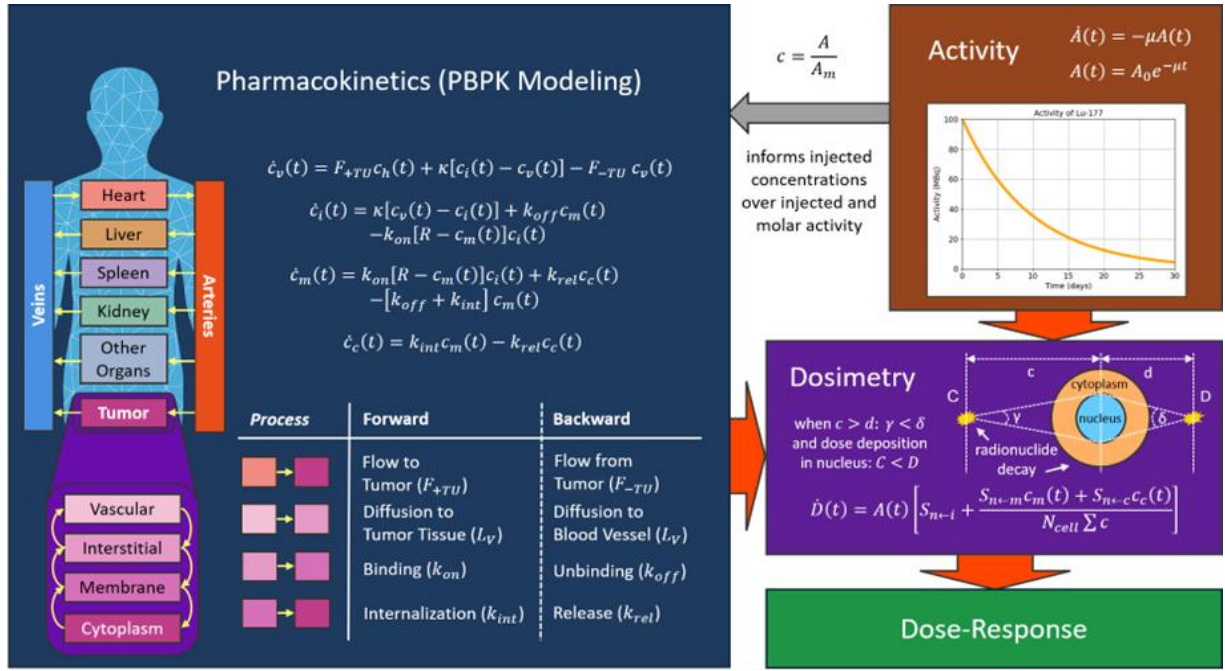

**Figure 5** Possible mathematical framework of a mechanistic theranostic digital twin (TDT). PBPK models can simulate compartmental radiopharmaceutical concentrations ($c_x$) at both the organ and cell levels, and they depend on kinetic parameters and target receptor concentration ($R$) [43,122]. Together with the physical activity ($A$), determined by the decay rate ($\lambda$), and dosimetry S values ($S_{n\leftarrow x}$), a measure of cellular dose deposition per decay, these concentrations can be used to derive an absorbed dose[45,62]. The corresponding dose–response relationships can then be calculated directly from the absorbed dose rate curve[73].

Mechanistic modeling of RPTs spans multiple domains: pharmacokinetics, nuclear and particle physics (for dosimetry), and cell/molecular biology (for dose–response modeling)[123]. Mathematical models are, by definition and by design, simplifications of reality[124]. Therefore, it is essential to extract the most relevant processes to capture the key quantitative relationships for the specific application[125]. For instance, while EBRT allows a relatively direct mapping from dose to survival, RPTs involve complex temporal dose distributions and dynamic biological repair processes that undermine the validity of simple LQ-based estimates, especially in the absence of a well-defined dose-rate response curve[78,126]. Incorporating intermediate modeling steps, such as DNA damage induction and repair, can correct this bias[127]. Whenever possible, models should be formulated with minimal complexity that still capture the essential behavior to avoid overfitting and ensure broader applicability.

## 5. Future Directions and Challenges

The next decade is expected to shift RPTs from protocol-centered delivery to a learning health system that relies on patient–specific digital twin models. These virtual avatars integrate multiscale Monte Carlo dosimetry, PBPK modeling, quantitative descriptors of the tumor microenvironment, and pharmacodynamic modeling. This integration enables the decomposition of conventional regimens into optimally sequenced micro-courses, tailored in real-time to individual biological responses. Within this framework, clinicians can continuously track the adaptive dynamics of malignant cell populations and iteratively refine key therapeutic variables, including radiopharmaceutical activity, emission type, and fractionation schemes between treatment cycles. This strategy aims to sustain cytotoxic pressure on tumors while proactively maintaining normal tissue exposure below toxicity thresholds. Preliminary studies have demonstrated that voxel-level dose distribution models, coupled with dynamic

PBPK simulations, can enable personalized administration of $^{177}$Lu and $^{225}$Ac-labeled agents[128]. Concurrent advancements in total-body PET/CT and SPECT/CT[129], graphics processing unit (GPU)-accelerated computation[130], and edge-based processing[131] are significantly reducing data acquisition and simulation times, thereby facilitating clinically actionable, adaptive replanning within inter-cycle intervals.

The next critical evolution in RPT is the development of biologically informed modeling[132]. By incorporating parameters such as tissue hypoxia, DNA repair capacity, cell cycle progression, and immune checkpoint expression into mechanistic response surfaces, digital twins can now extend beyond physical dosimetry to predict immunogenic cell death and systemic abscopal effects[133]. These predictions enable rational design of synergistic treatment schedules that integrate RPT with modalities such as external beam radiation[134], PARP inhibition[135], or immune checkpoint blockade[136], with the aim of maximizing therapeutic efficacy while minimizing cumulative toxicity[137].

However, substantial technical and regulatory challenges must be addressed. The lack of standardized formats for integrating dynamic PET, liquid biopsy, and radiobiological data hinders the development of unified models[138]. Furthermore, predictive accuracy remains limited by sparse clinical datasets, requiring probabilistic methods such as Bayesian inference for uncertainty quantification[139]. Whole-organ Monte Carlo simulations are computationally intensive, and real-time dosimetry depends on high-performance and edge computing infrastructure[140], which remains unevenly accessible[141]. In addition, data ownership remains ambiguous, and ethical frameworks for cross-border privacy and digital governance are lacking[142]. Biologically, treatment efficacy is limited by tumor heterogeneity and dynamic antigen expression, leading to potential target escape[143]. The supply of α-emitters such as $^{225}$Ac is limited by its dependence on scarce precursors like $^{229}$Th[144]. Clinically, heterogeneous data standards and imaging protocols across institutions impede interoperability[145]. Effective deployment requires integrated workflows across imaging archives, dosimetry systems, electronic health records, radiopharmaceutical calibration, and waste management. Therefore, a verifiable, transparent regulatory framework remains essential.

Digital twin technology marks a shift in RPT from standardized protocols to biologically adaptive, patient–specific strategies. Integrating real-time physiological data into treatment planning enables precise modulation of therapeutic parameters. Its development requires coordinated advances in computation, biomedicine, engineering, regulation, and ethics, offering a meaningful opportunity to improve efficacy and reduce toxicity in cancer care.

## 6. Summary

TDTs provide a computational framework to tailor RPT schedules to individual patients, also including in the context of combination treatments, with the goal of optimizing therapeutic outcomes and improving survival. The overarching aim of TDT research is to establish mathematical and computational models that account for the diverse ways in which treatments interact with patients. These models are fed with individualized data to simulate treatment responses.

Conceptually, the TDT consists of dosimetry and dose–response, or alternatively, pharmacokinetics and pharmacodynamics. Dosimetry incorporates pharmacokinetics and part of pharmacodynamics to assess the energy deposited in tumor sites, often using tools like PBPK modeling and Monte Carlo simulations. Dose–response captures the remaining pharmacodynamic component, evaluating radiation-induced tissue damage and its cellular effects. This component is often modeled using the LQ model.

To shift RPT to patient–specific strategies, collaboration between clinicians, engineers, and scientists is essential. Biologists, computer scientists, mathematicians, and physicists must develop models that are both accurate and practical, while clinicians are tasked with embracing innovation and adapting treatment paradigms. Together, these interdisciplinary efforts present a powerful opportunity to meaningfully improve cancer prognosis and patient care.

## Clinical Care Points

- Algorithms will be able to propose optimal RPTs treatment plans based on diagnostic data, including combination regimens with immunotherapies.
- It is recommended that preclinical studies be designed based on insights from mathematical modeling to yield the most relevant findings.